\documentclass[showpacs,aps,prl,twocolumn,superscriptaddress]{revtex4-1}
\usepackage{graphicx} 
\usepackage{dcolumn}
\usepackage{bm}
\usepackage{amssymb,amsmath}
\usepackage{epstopdf}
\usepackage{color}
\def\etal{$\it{et~al.}$}

\begin{document}
\title{Electrically controlled terahertz magneto-optical phenomena in continuous and patterned graphene}

\author{Jean-Marie Poumirol}
\email{Jean-Marie.Poumirol@unige.ch}
\affiliation{Department of Quantum Matter Physics, University of Geneva, CH-1211 Geneva 4, Switzerland}
\author{Peter Q. Liu}
\affiliation{Institute for Quantum Electronics, Department of Physics, ETH Zurich, CH-8093 Zurich, Switzerland}
\author{Tetiana M. Slipchenko}
\affiliation{Instituto de Ciencia de Materiales de Aragon and Departamento de Fisica de la Materia Condensada,
CSIC-Universidad de Zaragoza, E-50009, Zaragoza, Spain}
\author{Alexey Yu. Nikitin}
\affiliation{CIC nanoGUNE, E-20018, Donostia-San Sebasti\'{a}n, Spain}
\affiliation{IKERBASQUE, Basque Foundation for Science, 48011 Bilbao, Spain}
\author{Luis Martin-Moreno}
\affiliation{Instituto de Ciencia de Materiales de Aragon and Departamento de Fisica de la Materia Condensada,
CSIC-Universidad de Zaragoza, E-50009, Zaragoza, Spain}
\author{J\'{e}r\^{o}me Faist}
\affiliation{Institute for Quantum Electronics, Department of Physics, ETH Zurich, CH-8093 Zurich, Switzerland}
\author{Alexey. B. Kuzmenko}
\email{Alexey.Kuzmenko@unige.ch}
\affiliation{Department of Quantum Matter Physics, University of Geneva, CH-1211 Geneva 4, Switzerland}

\date{\today}

\begin{abstract}
\textbf{The magnetic circular dichroism and the Faraday rotation are the fundamental phenomena of great practical importance arising from the breaking of the time reversal symmetry by a magnetic field. In most materials the strength and the sign of these effects can be only controlled by the field value and its orientation. Furthermore, the terahertz range is lacking materials having the ability to affect the polarisation state of the light in a non-reciprocal manner. Here we demonstrate, using broadband terahertz magneto-electro-optical spectroscopy, that in graphene both the magnetic circular dichroism and the Faraday rotation can be modulated in intensity, tuned in frequency and, importantly, inverted using only electrostatic doping at a fixed magnetic field. In addition, we observe strong magneto-plasmonic resonances in a patterned array of graphene antidots, which potentially allows exploiting these magneto-optical phenomena in a broad THz range.}
\end{abstract}


\maketitle

\noindent \textbf{Introduction}

Terahertz waves are actively explored for fundamental research and applications \cite{PeiponenBook12,ZouaghiEJP13}. However, there is a lack of materials and technology allowing the fabrication of efficient and broadly tunable passive terahertz devices. Graphene, where the doped charge carriers interact strongly with the terahertz radiation \cite{HorngPRB11,JuNN11,SensaleNatComm12,RenNL12,WitowskiPRB10,CrasseeNP11,YuPRB16} and the doping is electrostatically tunable \cite{NovoselovScience04}, is promising to tackle this problem \cite{VakilScience11,SounasAPL13,TamagnoneNP14}. Moreover, graphene-light interaction is greatly enhanced if graphene plasmons \cite{WunschNJP06,HwangPRB06,JablanPRB09,BonnacorsoNatPhot10,KoppensNL11,ChenNature12,FeiNature12,GrigorenkoNPh12,LowACSNano14} are excited by patterning \cite{JuNN11,YanNL12,LiuOptica15}. An additional advantage of graphene is the small cyclotron mass making the Drude response highly sensitive to the magnetic field \cite{WitowskiPRB10,CrasseeNP11}. While the mass is also small in conventional two-dimensional electron gases (2DEGs) \cite{StormerSSC79}, the fundamental difference of graphene is that the charge carriers are Dirac fermions, where the cyclotron mass depends on doping \cite{AndoJPSJ02} and that the ambipolar gating is possible \cite{NovoselovScience04}. Moreover, while cryogenic temperatures are needed to observe magneto-optical (MO) effects in 2DEGs, in graphene they show up already at room temperature \cite{CrasseePRB11}. Unfortunately, not all the mentioned graphene benefits were explored so far in a single experiment.

Here we combine magneto-optical THz spectroscopy with electrostatic gating and graphene patterning. This allows us to observe several new phenomena, including an electrostatic control of the THz cyclotron frequency, a purely electrostatic inversion of the magnetic circular dichroism and the Faraday rotation and strong gate-tunable magneto-plasmonic resonances in periodic antidot arrays. We argue that these effects potentially allow a broad spectral tunability and enable novel functionalities in graphene-based passive terahertz devices.\\

\noindent \textbf{Results}

\noindent \textbf{Continuous graphene}


\begin{figure*}
\includegraphics[width=18cm]{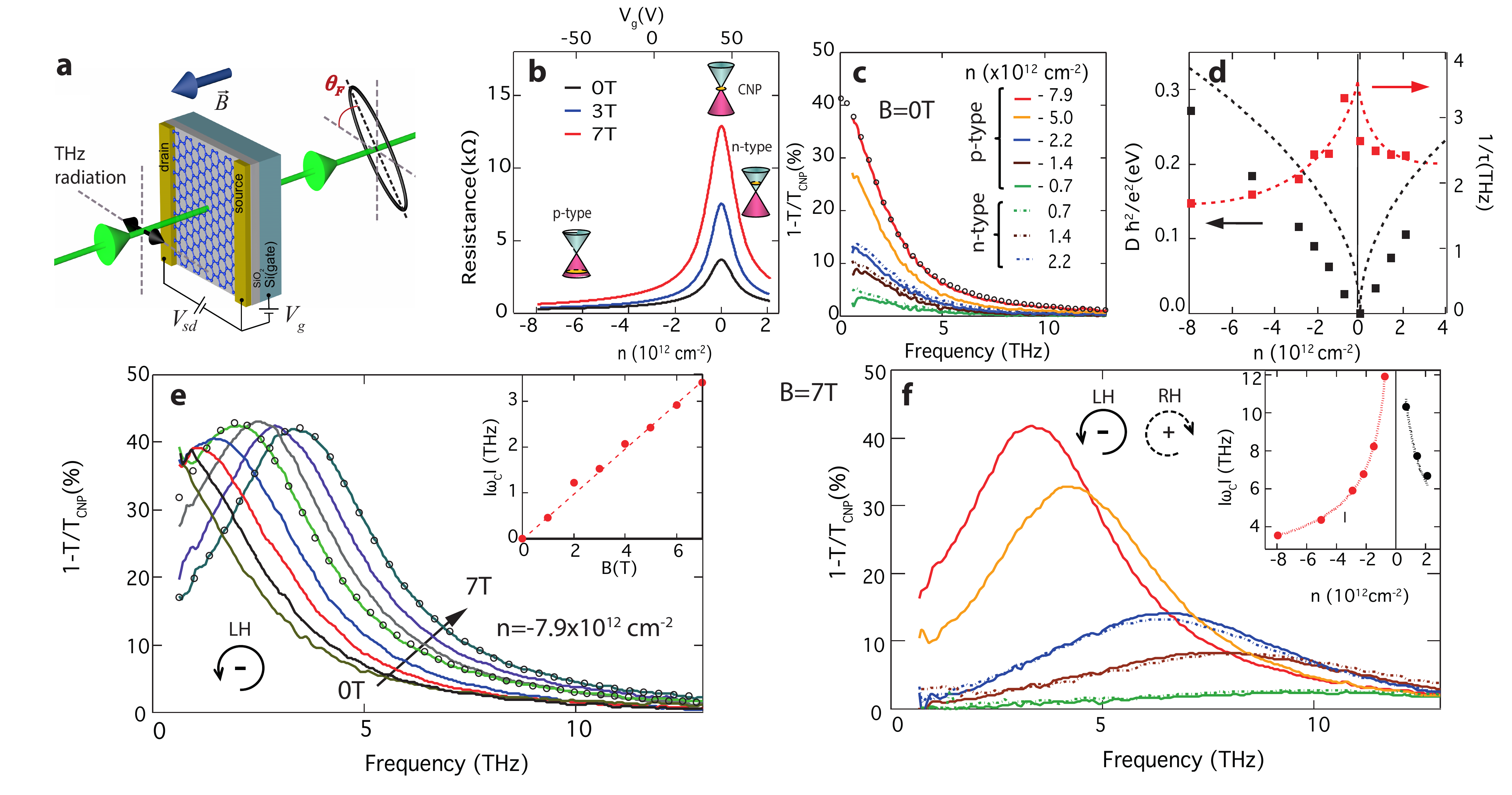}
\caption{\label{FigCont} \textbf{Doping dependent magneto-resistance and magneto-optical extinction spectra of continuous graphene.} \textbf{a}, schematic representation of a terahertz g-FET device and the optical experiment.  \textbf{b}, Two-terminal source-drain resistance at $B$ = 0, 3 and 7 T as a function of the gate-induced doping. \textbf{c}, Extinction spectra at $B$=0 T for various doping levels and polarities. The open circles represent the best Drude fit for $n=-7.9\times10^{12}$ cm$^{-2}$. \textbf{d}, Doping dependence of the Drude weight relative to the CNP (black symbols) and the scattering rate (red symbols). The black line is the theoretical prediction for Drude weight of Dirac fermions for $v_\text{F}=10^6$ m s$^{-1}$.  The red line is the guide to the eye. \textbf{e},  The extinction (for the LH circular polarization) at $n=-7.9\times10^{12}$ cm$^{-2}$ for different values of magnetic field from 0 T to 7 T with the step of 1 T. The open circles represent the best Drude fits at 4 and 7 T. The inset presents the field dependence of the CR frequency (circles) and the linear fit (dashed line). \textbf{f}, The extinction spectra at 7 T as a function of doping. The color legend is the same as in panel \textbf{c}. The spectra are shown for the LH/RH circular polarizations for the p-/n-doped regimes respectively, \emph{i.e. }for polarizations exhibiting the cyclotron resonance. The inset presents the doping dependence of the experimental cyclotron frequency (circles) and a fit using a Dirac-fermion model as described in the text (dashed lines). All measurements are done at $T$ = 250 K. }
\end{figure*}

\begin{figure}
\includegraphics[width=8cm]{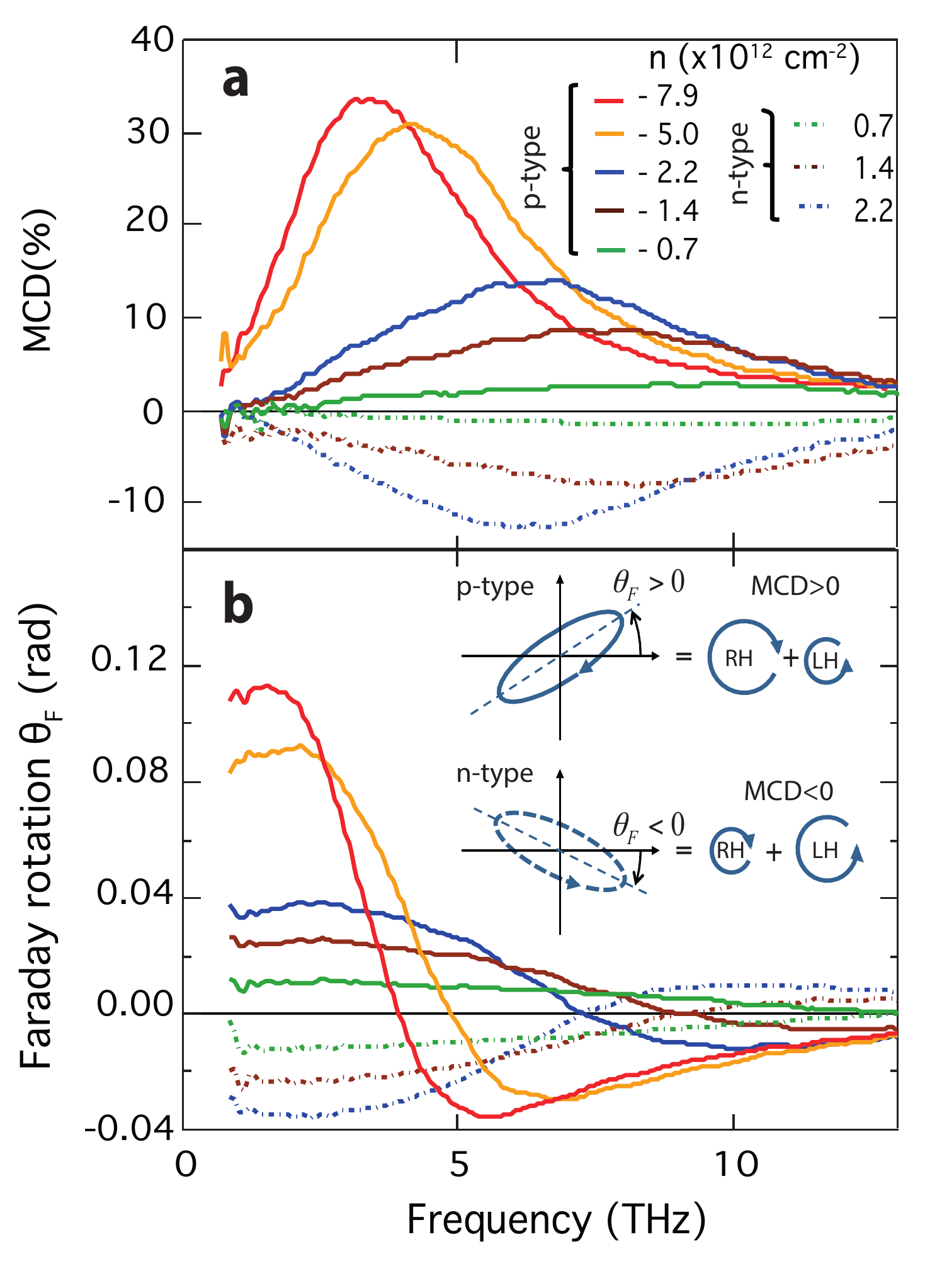}
\caption{\label{FigInv} \textbf{Electrostatic sign inversion of magnetic circular dichroism and Faraday rotation in continuous graphene.} \textbf{a} and \textbf{b}, spectra of MCD=$(T_{-}-T_{+})/T_{\text{CNP}}$ and FR $\theta_{\text{F}}$ at different doping levels at 7 T. The inset compares the polarization ellipses of the transmitted radiation and illustrates the opposite signs of the MCD and the FR  for the p- and n-type doping. All measurements are done at $T$ = 250 K. }
\end{figure}

We first discuss the intrinsic MO terahertz effects in continuous (unpatterned) graphene. A basic device used in our experiments is a large-area CVD-graphene field-effect transistor (g-FET) sketched in Fig.~\ref{FigCont}a. It allowed us to combine ambipolar control of doping $n$ (positive for electrons and negative for holes) with magneto-transport and magneto-optical measurements in a perpendicular magnetic field $B$ (positive towards the light source). The THz transmission $T(\omega)$ and the Faraday rotation spectra $\theta_{\text{F}}(\omega)$ were obtained, while graphene was illuminated by a linearly polarized light with the angular frequency $\omega$. The Fabry-Perot interference in the substrate was suppressed intentionally to simplify the spectra analysis. From these measurements we extracted the transmission spectra for the right-hand(RH)/left-hand (LH) circular polarizations, $T_{\pm}(\omega)$ \cite{LevalloisRSI15}, which fully characterizes the MO response of the device.

The source-drain resistance as a function of the gate voltage $V_{\text{g}}$ in all our devices has a characteristic peaked shape (Fig.~\ref{FigCont}b) with the charge neutrality point (CNP) at a positive voltage, indicating a residual p-type doping.  The curves show a strong positive magnetoresistance and are electron-hole symmetric, demonstrating the high quality of our samples. The charge-carrier mobility $\mu \approx 3,500$ cm$^2$V$^{-1}$s$^{-1}$ and the doping inhomogeneity $\delta n \approx 5\times 10^{11}$cm$^{-2}$ are deduced using a standard analysis of the transport curves \cite{KimAPL09}. Fig.~ \ref{FigCont}c shows the zero-field extinction spectra $1 - T(\omega)/T_{\mbox{\tiny\text{CNP}}}(\omega)$, which are representative of the optical absorption in graphene (as specified in Supplementary Note 2 and shown in Supplementary Figure 2). The strong exctinction increase at low frequencies is due to the Drude absorption by the doped charges \cite{HorngPRB11,JuNN11,SensaleNatComm12,RenNL12,YuPRB16}. The fact that almost half of terahertz photons are stopped by gate-injected carriers in an atomic monolayer demonstrates the remarkable efficiency of the Drude response. Importantly, the extinction curves at the matching p- and n-type doping levels are similar, which is consistent with the symmetric shape of the transport curves.

Optical conductivity is described almost perfectly (as shown by the open circles in Fig.~\ref{FigCont}c) using a semiclassical Drude model:
\begin{equation}\label{Drude}
\sigma_{\pm} (\omega)=\frac{D}{\pi}\frac{i}{\omega\mp\omega_\text{c}+i/ \tau},
\end{equation}
\noindent where $D$ is the Drude weight, $\omega_{\text{c}}$ is the cyclotron frequency (positive/negative for n-/p-type doping), $\tau$ is the scattering time, and +/- refer to RH/LH circular polarizations. A relation between $\sigma_{\pm}$ and $T_{\pm}$ that takes the substrate into account, is given in Supplementary Note 1 and illustrated in Supplementary Figure 1.

Before we analyse the dependence of the Drude parameters on the doping and the magnetic field, we stress that the Dirac-fermion theory (in the semiclassical limit) predicts \cite{AndoJPSJ02} that
 \begin{equation}\label{DrudeWeight}
D(n)=\frac{e^2v_{\text{F}}}{\hbar}\sqrt{\pi |n|},
\end{equation}
\noindent and
\begin{equation}\label{OmegaC}
\omega_{\text{c}}(n,B)= \frac{eB v_{\text{F}}}{\hbar}  \frac{\text{sign}(n)}{\sqrt{\pi |n|}}.
\end{equation}
\noindent where $v_{\text{F}}$ is the Fermi velocity, $\hbar$ the reduced Planck's constant and $e$ the (positive) elementary charge. This behaviour is different from the case of conventional 2DEGs \cite{StormerSSC79}, where the Drude weight is proportional to $n$, while the cyclotron frequency is $n$-independent. Notably, the relation (\ref{OmegaC}) was not tested so far since in all terahertz spectroscopy measurements the electrostatic gating was done separately from applying a magnetic field.

The Drude weight and the scattering rate $1/\tau$ extracted by fitting the zero-field extinction curves (where $\omega_{\text{c}} = 0$) are strongly doping dependent (Fig.~\ref{FigCont}d). A theoretical dependence (dashed line) for a typical value of $v_{\text{F}}= 10^{6}$ m/s agrees reasonably well with our experiment, with the experimental values being about 20\%  lower, in a qualitative agreement with a previous report \cite{HorngPRB11}. We did not observe any significant change of the Drude weight with the magnetic field. As in a recent report \cite{YuPRB16}, the scattering rate decreases as the doping increases for both charge polarities. This effect, which is beneficial for MO applications, is consistent with either a doping-induced screening of charged impurities \cite{AndoJPSJ06,HwangPRL07,AdamPNAS07} or with the resonant scattering \cite{NiNL10}.

In a magnetic field the terahertz spectra are dominated by the cyclotron resonance (CR) (Figs. \ref{FigCont}e and \ref{FigCont}f). The extinction curves for a strongly p-type doped sample ($n = -7.9\times 10^{12}$ cm$^{-2}$) for the LH circular polarization, where the CR is observed in this case, are shown in Fig.\ref{FigCont}e for different magnetic field values up to 7 T. The Drude fit works equally well in magnetic field, as shown by open circles for 4 and 7 T. The cyclotron frequency shifts linearly with $B$ (inset of Fig. \ref{FigCont}e) and a record extinction of 43\% is reached at 5 T at a frequency as high as 2.5 THz. Furthermore, Fig. \ref{FigCont}f reveals that the cyclotron frequency strongly increases (up to 10 THz) with the reduction of the charge concentration for both doping regimes. This is a spectacular fingerprint of Dirac fermions, as theory indeed predicts $\omega_{\text{c}}$ to be doping dependent, according to Eq. (\ref{OmegaC}). This relation fits the experimental data very well (inset of  Fig.~\ref{FigCont}f), where similar Fermi velocity values for p-type and n-type doping regimes are found ($1.01 \pm 0.02$ and $0.92 \pm 0.05\times 10^{6}$  m.s$^{-1}$ respectively).

Fig.~\ref{FigInv} shows the magnetic circular dichroism (defined as the difference between the extinction coefficients for the RH and LH circular polarizations), and the FR spectra  at 7 T for different doping values. Both quantities can be non-zero only if the time reversal symmetry is broken and are the key parameters for the non-reciprocal light manipulation. We note that the high values of the MCD and the FR (in the present case 35\% and 0.11 rad respectively) demonstrate the potential of graphene as a MO material in this frequency range. Furthermore, Fig.~\ref{FigInv} reveals that in addition to tuning the frequency and modulating the intensity of the magneto-extinction, the electrostatic doping also allows the inversion of the magneto-circular dichroism and the Faraday rotation at a fixed magnetic field. Indeed, in both cases the spectra corresponding to equal absolute carrier concentrations but opposite doping types show a high degree of symmetry with respect to zero. This fully agrees with the Dirac-fermion theory predicting the inversion of the cyclotron frequency (Eq. \ref{OmegaC}) at a constant Drude weight (Eq. \ref{DrudeWeight}) as $n$ changes sign.  The inversion of the MCD is thus due to the fact that the polarization where the cyclotron resonance is observed is defined by the doping type. This new effect that enables novel applications (as discussed below) is thus directly related to the unique possibility of ambipolar doping control in graphene \cite{NovoselovScience04}.

\noindent \textbf{Patterned graphene}

\begin{figure*}
\includegraphics[width=18cm]{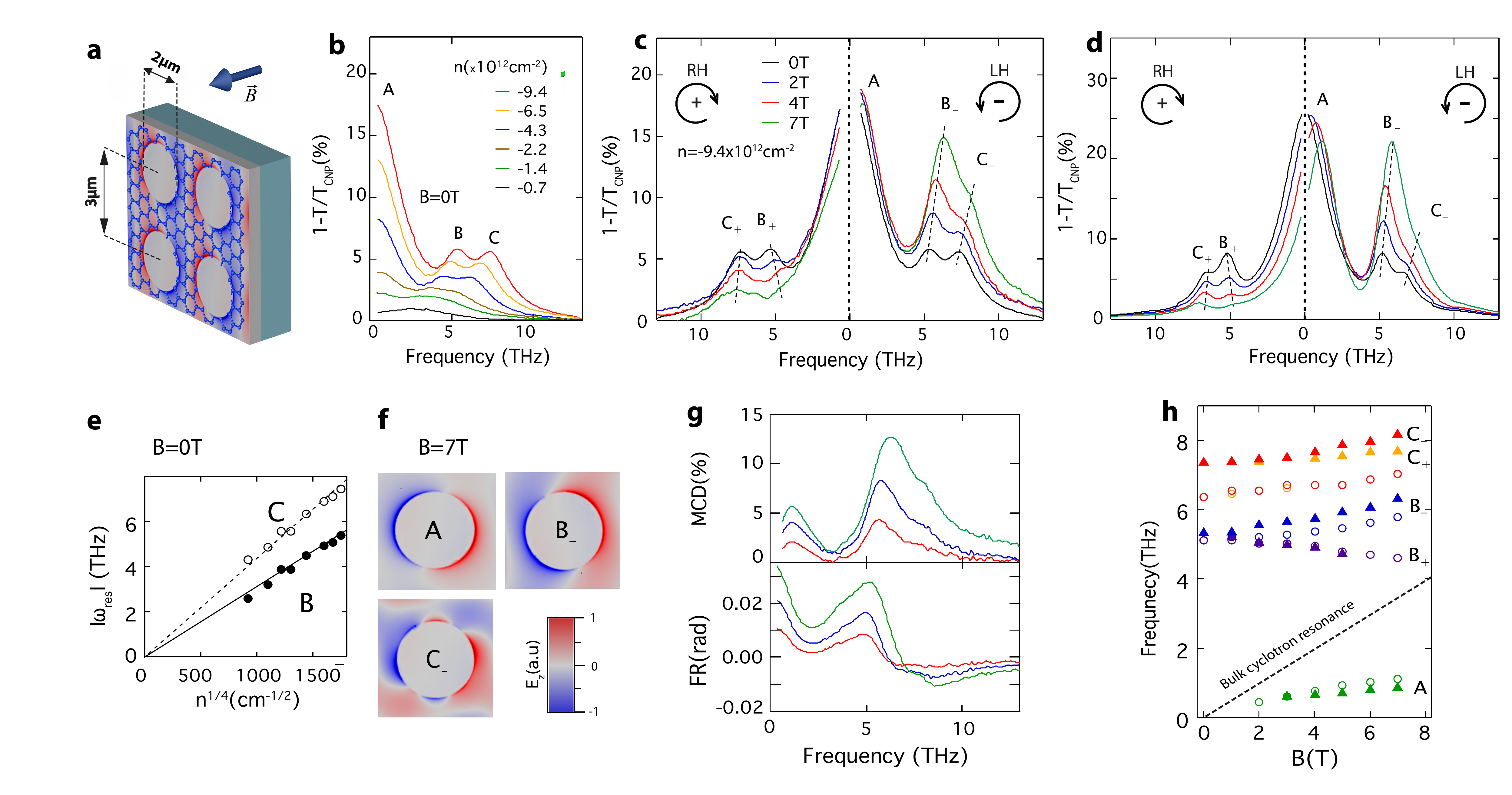}
\caption{\label{FigPatt} \textbf{Electro-magneto-optical terahertz experiment and simulations in antidot-patterned graphene.} \textbf{a}, Schematic illustration of the patterned antidot array. Electrical contacts and terahertz light, are the same as in Fig.\ref{FigCont}a, and are not shown. The superimposed color image exemplifies a distribution of the perpendicular AC electric field, when a magneto-plasmon mode ($B_{-}$) is excited. \textbf{b}, Experimental extinction spectra at $B$=0 T for various carrier concentrations (p-doping). Symbols A, B and C indicate the resonant peaks. \textbf{c}, Experimental MO extinction spectra for RH and LH polarizations at $n = -9.3\times 10^{12}$ cm$^{-2}$ at selected magnetic fields. The black dashed lines are guides to the eye to follow the evolution of peaks B and C. \textbf{d}, Finite-element simulation of the experimental data at the same doping level and the magnetic field values as in panel \textbf{c}. As in panel \textbf{c}, the black dashed lines follow peaks B and C. \textbf{e}, Doping dependence of the Bragg peaks B and C extracted from \textbf{b}.  \textbf{f}, Distribution of AC electric field $E_{z}$ corresponding to the modes A$_{-}$, B$_{-}$ and C$_{-}$ at 7 T as shown in \textbf{d} . \textbf{g}, Spectra of MCD and FR corresponding to panel \textbf{c}. \textbf{h}, Magnetic-field dependence of experimental (solid triangles) and simulated (open circles) magnetoplasmon frequencies corresponding to panels \textbf{c} and \textbf{d}. The black dashed line indicates the bulk cyclotron resonance $\omega_{\text{c}}(B)$. All measurements are done at $T$ = 250 K. }
\end{figure*}

As demonstrated above, the non-reciprocal MO effects in unpatterned graphene are especially strong at low frequencies (below 5-7 THz). However, in the higher-THz region the Drude response is less efficient. In order to broaden further the tuning range of the magneto-teraherz response, one can use the new physical properties arising from the excitation of magneto-plasmons (\emph{i.e.} collective charge oscillations in external magnetic field) in graphene nanostructures \cite{CrasseeNL12,YanNL12,WangPRB12,TymchenkoACSNano13,PetkovicPRL13}. To this end we patterned graphene with a periodic antidot array (Fig.~\ref{FigPatt}a) that acts as a two-dimensional plasmonic crystal \cite{LiuOptica15}. The transport curves (presented in the Supplementary Fig. 3) are similar to the ones in continuous graphene (Fig.~\ref{FigCont}b) indicating that patterning does not strongly degrade the charge mobility. In order to better understand the magnetic field dependence we first show in Fig.~ \ref{FigPatt}b the zero-field extinction spectra of patterned graphene with a period of 3 $\mu$m, and a hole diameter of 2 $\mu$m that were studied previously by us \cite{LiuOptica15}. They feature a Drude-like peak (A) as in continuous graphene and two additional peaks (B and C) due to Bragg scattering of graphene plasmons on the periodic structure.  Our previous study revealed that peak B is mostly due to the (0,1)-order Bragg reflection, while peak C has a mixed (1,1) - (0,2) character. Importantly, they both demonstrate a $\omega_{\mbox{\tiny res}}\sim n^{1/4}$ dependence typical for Dirac fermions (Fig.~\ref{FigPatt}e) \cite{WunschNJP06,HwangPRB06,JuNN11} that allows controlling the resonance frequency by the gate voltage, in addition to tuning it by the antidot dimensions and spacing \cite{LiuOptica15}.

A comparison of Fig.\ref{FigPatt}b and Fig.\ref{FigCont}c reveals that the extinction in patterned graphene is in fact lower than or comparable to the extinction in continuous graphene at the same frequencies, even close to plasmonic resonances. A certain reduction of the extinction is because of the removal of graphene by patterning and the corresponding reduction of the filling factor (about 40\% in our case). Secondly, this is a consequence of a relatively low mobility, which broadens and reduces both the plasmon peak and the Drude peak. Therefore, in patterned samples with a high enough mobility, where the peaks are expected to be sharper and higher, the extinction should be higher than in continuous graphene.

The magnetic field affects strongly the extinction curves (Fig. \ref{FigPatt}c) and shifts the peak frequencies (solid symbols in Fig. \ref{FigPatt}h). Peak A shifts upwards in the LH polarization while keeping the intensity constant. Although this resembles the behaviour of the CR peak in continuous graphene (Fig. ~\ref{FigCont}e),  the frequency of peak A is much lower than the bulk CR for the same values of $n$ and $B$ and shows a sublinear field dependence. A similar excitation was observed in 2DEGs \cite{KernPRL91} and identified as a cyclotron mode at low magnetic fields crossing over to a so-called edge magnetoplasmon circulating around the holes as the field is increased. Bragg magnetoplasmon peaks B and C exhibit a dramatic field dependence and a totally different behaviour for the RH and LH circular polarizations (where we denote them B$_{+}$, C$_{+}$ and B$_{-}$, C$_{-}$ respectively). Most notable is a strong increase/decrease of their intensity for the LH/RH polarization, accompanied by field-induced shifts as indicated by the dashed lines in Fig. \ref{FigPatt}c. The polarization frequency splitting is most prominent for peak B, where opposite shifts are observed in the two polarizations. Associated to these spectacular magnetoplasmonic effects are the resonant MCD and FR structures (Fig.~\ref{FigPatt}g), which are present at much higher frequencies than resonances in continuous graphene (Fig.~\ref{FigInv}), confirming that patterning indeed extends significantly the spectral range of the strong MO activity.

These results agree qualitatively with low-temperature experiments \cite{KernPRL91,BollwegPRL96,HochgrafePRB99} and theory \cite{MikhailovPRB95,MikhailovPRB96} in conventional 2DEG antidot arrays. However, it is imperative to quantitatively understand the present spectra in order to predict the performance of graphene THz devices. To this end we performed a finite-element electromagnetic simulation of our system, where the Drude weight $D$ and the cyclotron frequency $\omega_{c}$ were determined using Eqs. (\ref{DrudeWeight}) and (\ref{OmegaC}), while the scattering time $\tau$ was set to 0.1 ps in order to match the observed width of the magnetoplasmon peaks. The simulated extinction spectra for various magnetic field values are shown in Fig.~\ref{FigPatt}d and the corresponding mode frequencies are shown as open circles in Fig.~\ref{FigPatt}h One can notice an excellent agreement with the experimental curves (Fig.~\ref{FigPatt}c), in terms of the spectral shapes and the evolution of the peak positions and intensities as a function of $B$. In particular, the simulation reproduces perfectly the sublinear field dependence of peak A and the opposite field-induced shifts of the B$_{-}$ and B$_{+}$ peaks.

To give a further physics insight into the origin of the observed MO resonances, we show in Fig.~\ref{FigPatt}f the simulated distributions of the AC electric field $E_{z}$ perpendicular to the sample for the A, B$_{-}$ and C$_{-}$ modes at 7 T. The electric field for the mode A is strongly localized near the antidot edge, corroborating its prior identification as an edge magnetoplasmon \cite{MikhailovPRB95,MikhailovPRB96,HochgrafePRB99} where the cyclotron orbits skip around the antidots. In contrast, the Bragg modes B$_{-}$ and C$_{-}$ are characterized by peculiar field distributions in the entire unit cell. At the edge, mode B$_{-}$ has two $E_{z}$ nodes that arise from the coupling between the dipolar plasmon and the CR. Mode C$_{-}$ exhibits six nodes, originating from the mentioned above mixed-order Bragg reflection character. Because of this, mode C may be more sensitive to a possible variation of the graphene doping near the edge, not taken into account in the simulation, which could be the reason for some discrepancy (about 1 THz) between the experimental and the simulated frequencies. Nevertheless, overall our simulations based on the Dirac-fermion theory prove to be quantitatively accurate also in patterned graphene.

\noindent \textbf{Discussion}

Our results indicate that combining terahertz spectroscopy, electrostatic gating, magnetic field and pattering reveals interesting properties of the Dirac fermions, such as a broadband electrostatic control of the cyclotron frequency and relativistic magnetoplasmons, and give rise to qualitatively new phenomena, such as a purely electrostatic control of the magnetic effects (MCD and FR). Importantly, the experiments were done close to room temperature, in a striking contrast to the existing 2DEG studies \cite{KernPRL91,BollwegPRL96,HochgrafePRB99}, where the samples have to be cooled below 4 K. This suggests the fundamental possibility of graphene-based terahertz non-reciprocal devices exploiting these effects. Examples of such devices are optical modulators, polarization converters, isolators and circulators, which are essential elements lacking in the present and future terahertz applications (life sciences, material characterization, telecommunications, homeland security ...). As the strength of MO effects depends critically on the charge mobility, improving this parameter is essential to achieve a higher MO performance. The value of $\mu\approx 3,500$ cm$^2$V$^{-1}$s$^{-1}$ in our present g-FETs is largely limited by the substrate (SiO$_{2}$/Si), and high magnetic fields (up to 7 T) are needed to attain reasonably good device specifications \cite{TamagnoneNC16}. On the other hand, much higher mobility values  \cite{DeanNN10} and much longer optical scattering times \cite{WoessnerNM15,DaiNN15} can be achieved in graphene on boron nitride.  Notably, a mobility of $\sim 100,000$  cm$^2$V$^{-1}$s$^{-1}$ at room temperature was recently obtained in h-BN-encapsulated CVD graphene using a dry-transfer technique \cite{BanszerusSA15}. Although at present the small size of graphene-BN samples make MO THz experiments difficult, using CVD technique in combination with h-BN encapsulation \cite{BanszerusSA15} is highly promising for future large-scale applications of high-mobility graphene. This would allow using cheap permanent magnets making the teragertz MO applications commercially relevant.

As we demonstrated, the unusual properties of Dirac fermions allow a wide tuning of the working (cyclotron or magneto-plasmon) frequency, potentially in the entire THz range, by the doping, magnetic field and patterning parameters. Moreover, the possibility to invert the magnetic circular dichroism and the Faraday rotation electrically opens the door to completely new functionalities, such as electrically switchable isolators, and to combining several basic functions in a single device.

Importantly, in the present case the magneto-optical effects occur in an atomically thin layer, therefore the device thickness is limited only by the substrate and external optical elements (polarizers, waveplates, magnets). This potentially allows a miniaturization of the graphene-based elements, in a stark contrast with conventional ones, where light has to travel a centimeter-scale distance. The scope of the MO terahertz applications of graphene is certainly not limited to free-space wave propagation, as in the present experiments, and we anticipate that its remarkable properties will be employed in miniaturized chip- and THz waveguide-integrated \cite{VitielloJAP11} devices.

\noindent \textbf{Methods}

\emph{Experiment. } Large-area monolayer CVD graphene was grown on copper and transferred to a weakly doped Si substrate with a 300 nm layer of SiO$_2$ that served as a back-gate dielectric. Au/Ti electrodes were deposited outside the illuminated region (5$\times$5 mm$^2$) allowing transport measurements and back-gate control simultaneously with magneto-optical studies. The carrier concentration was calculated using the formula $n=\alpha(V_{\text{g}} - V_{\text{g},\mbox{\tiny \text{CNP}}})$, where $\alpha = 7.2\times 10^{10}$ cm$^{-2}$V$^{-1}$ is the gate capacitance and $V_{\text{g},\mbox{\tiny \text{CNP}}}$ is the gate voltage corresponding to the maximum resistivity. A perpendicular magnetic field up to 7 T was applied using a superconducting split-coil system. The sample was illuminated by linearly polarized light and the terahertz transmission and Faraday-rotation spectra using the technique presented in Ref.\cite{LevalloisRSI15}, were measured  close to the room temperature (250 K) as a function of the carrier concentration and magnetic field with the help of a Fourier transform spectrometer equipped with a bolometer detector. In order to simplify the data analysis, the Fabry-Perot interference in the Si substrate (0.5 mm thick in our samples) was suppressed by reducing the optical resolution to 0.2 THz. By applying magneto-optical  Kramers-Kronig analysis \cite{LevalloisRSI15} to these data (shown in Supplementary Figure 4), we derived the transmission spectra for right-handed (RH) and left-handed (LH) circular polarizations without the need of physically generating circularly polarized waves.

\emph{Simulations. }Finite-element electromagnetic simulations of antidot arrays were performed using COMSOL software. The graphene layer was modeled as a surface current in the boundary conditions. In order to achieve convergence, the mesh element size in the vicinity of graphene was much smaller than the magneto-plasmon wavelength.

\emph{Data availability. } The data that support the findings of this study are available from the corresponding author upon request.

\noindent \textbf{Acknowledgements}

This research was supported  by the EU Graphene Flagship (Contract No. CNECT-ICT-604391 and 696656) and by the Swiss National Science Foundation (Grant No. 200020-156615). A.Y.N. acknowledges support from the Spanish Ministry of Economy and Competitiveness (national project MAT2014-53432-C5-4-R). J. F. acknowledges the ERC Grant "MUSIC". We thank M. Tran for valuable help and M. Tamagnone for useful discussions.

\noindent \textbf{Author contributions}

J.M.P. planned and performed all experiments, analysed the results and wrote the manuscript. P.Q.L. and J.F. fabricated g-FETs with continuous and patterned graphene, did transport measurements and discussed the results, T.M.S., A.Y.N. and L.M.M performed numerical calculations and discussed the results. A.B.K. planned all experiments, analysed the results and wrote the manuscript.

The authors declare no competing financial interests.

\end{document}


\title{Supplementary information: Electrically controlled terahertz magneto-optical phenomena in continuous and patterned graphene}

\author{Jean-Marie Poumirol}
\affiliation{Department of Quantum Matter Physics, University of Geneva, CH-1211 Geneva 4, Switzerland}
\author{Peter Q. Liu}
\affiliation{Institute for Quantum Electronics, Department of Physics, ETH Zurich, CH-8093 Zurich, Switzerland}
\author{Tetiana M. Slipchenko}
\affiliation{Instituto de Ciencia de Materiales de Aragon and Departamento de Fisica de la Materia Condensada,
CSIC-Universidad de Zaragoza, E-50009, Zaragoza, Spain}
\author{Alexey Yu. Nikitin}
\affiliation{CIC nanoGUNE, E-20018, Donostia-San SebastiÃ¡n, Spain}
\affiliation{IKERBASQUE, Basque Foundation for Science, 48011 Bilbao, Spain}
\author{Luis Martin-Moreno}
\affiliation{Instituto de Ciencia de Materiales de Aragon and Departamento de Fisica de la Materia Condensada,
CSIC-Universidad de Zaragoza, E-50009, Zaragoza, Spain}
\author{J\'{e}r\^{o}me Faist}
\affiliation{Institute for Quantum Electronics, Department of Physics, ETH Zurich, CH-8093 Zurich, Switzerland}
\author{Alexey. B. Kuzmenko}
\affiliation{Department of Quantum Matter Physics, University of Geneva, CH-1211 Geneva 4, Switzerland}
\begin{abstract}
\end{abstract}


\begin{figure*}
\includegraphics[width=8cm]{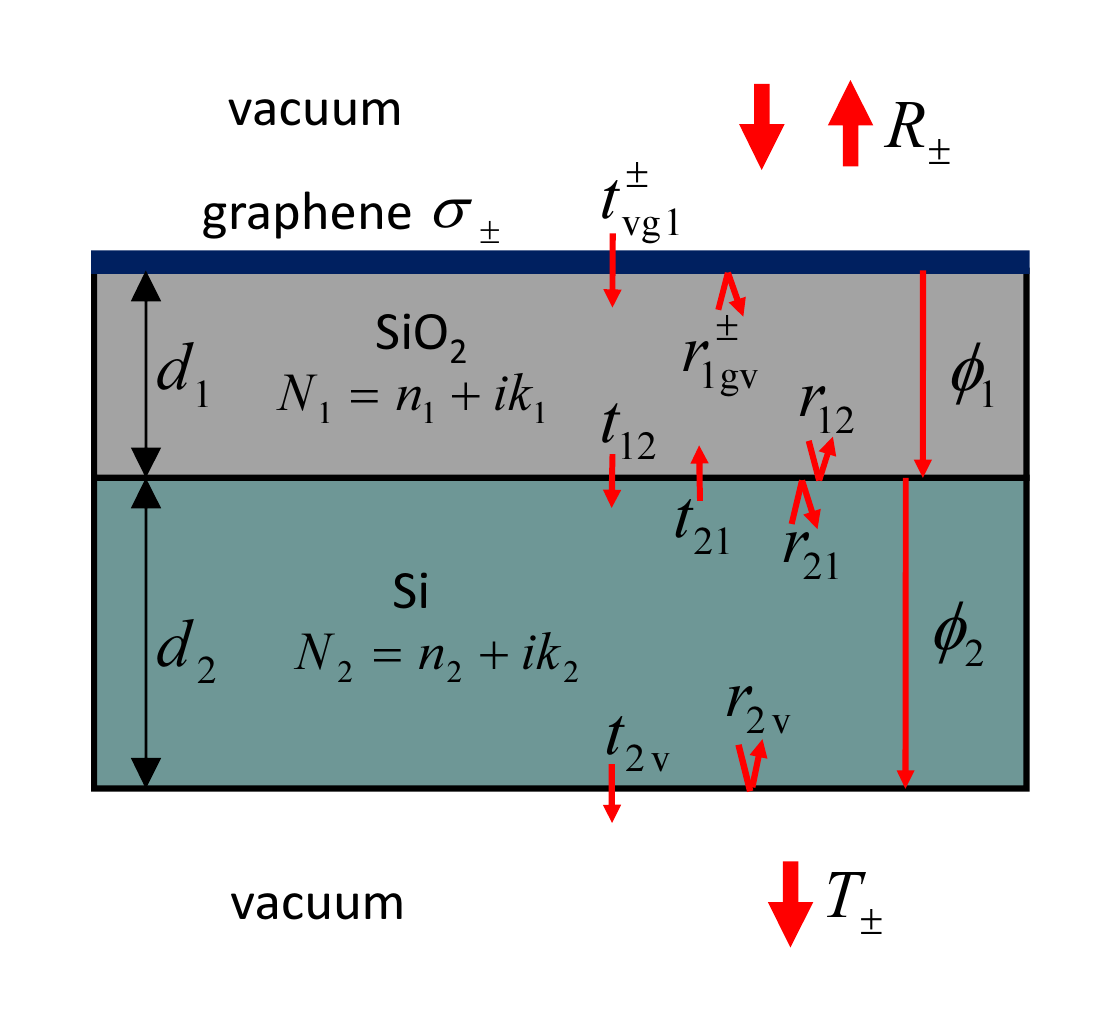}
\caption{\label{FigSetup} \textbf{Supplementary Figure 1.} Magneto-optical model of g-FET and description of the Fresnel coefficients used to compute the total magneto-optical transmission.}
\end{figure*}


\begin{figure*}[h]
\includegraphics[width=10cm]{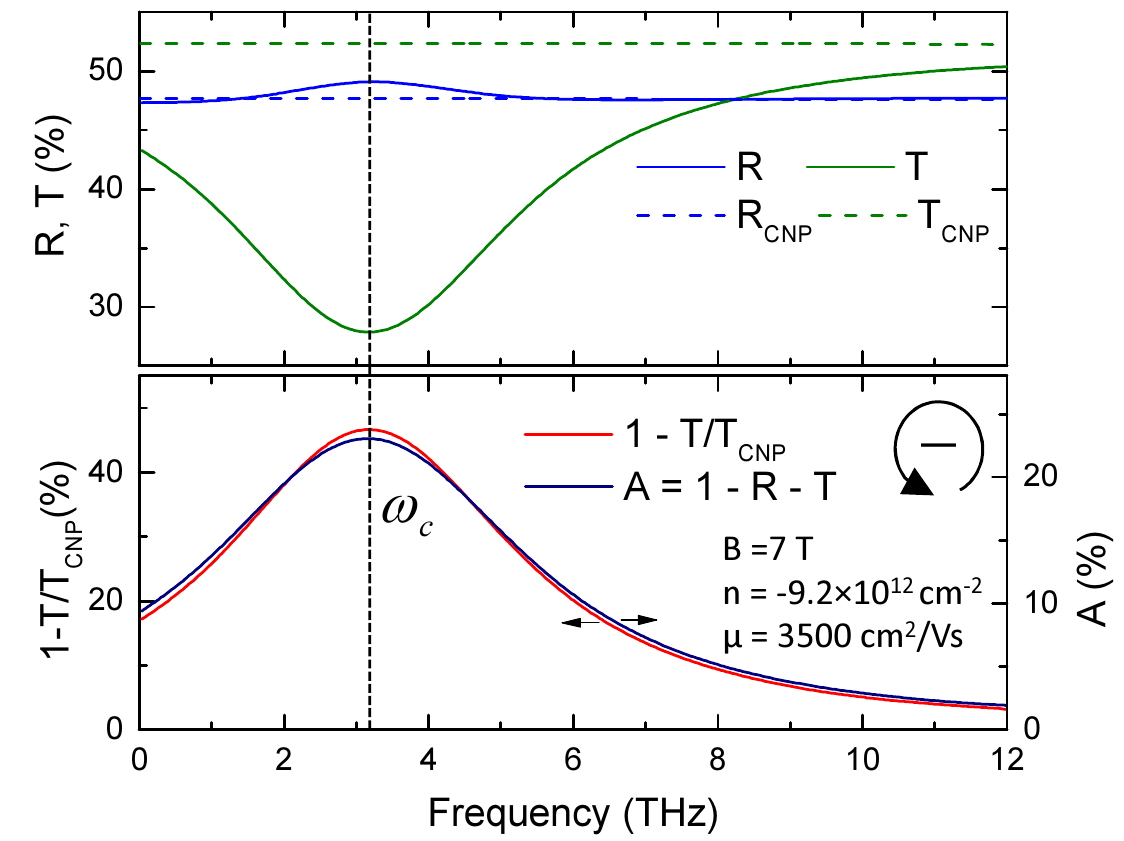}
\caption{\label{FigSpectra} \textbf{Supplementary Figure 2.} Top panel: model spectra of reflectivity (green), transmission (blue) of a g-FET at $B$ = 7 T with $n=-9.2\times 10^{12}$ cm$^{-2}$ (solid lines) and at the charge-neutrality point (dashed lines): The mobility $\mu$=3,500 cm$^2$V$^{-1}$s$^{-1}$, which is close to the experimental value. Bottom panel: the corresponding spectra of extinction $1 - T/T_{\text{CNP}}$ and absorption $A=1 - R - T$.}
\end{figure*}

\begin{figure*}[h]
\includegraphics[width=8cm]{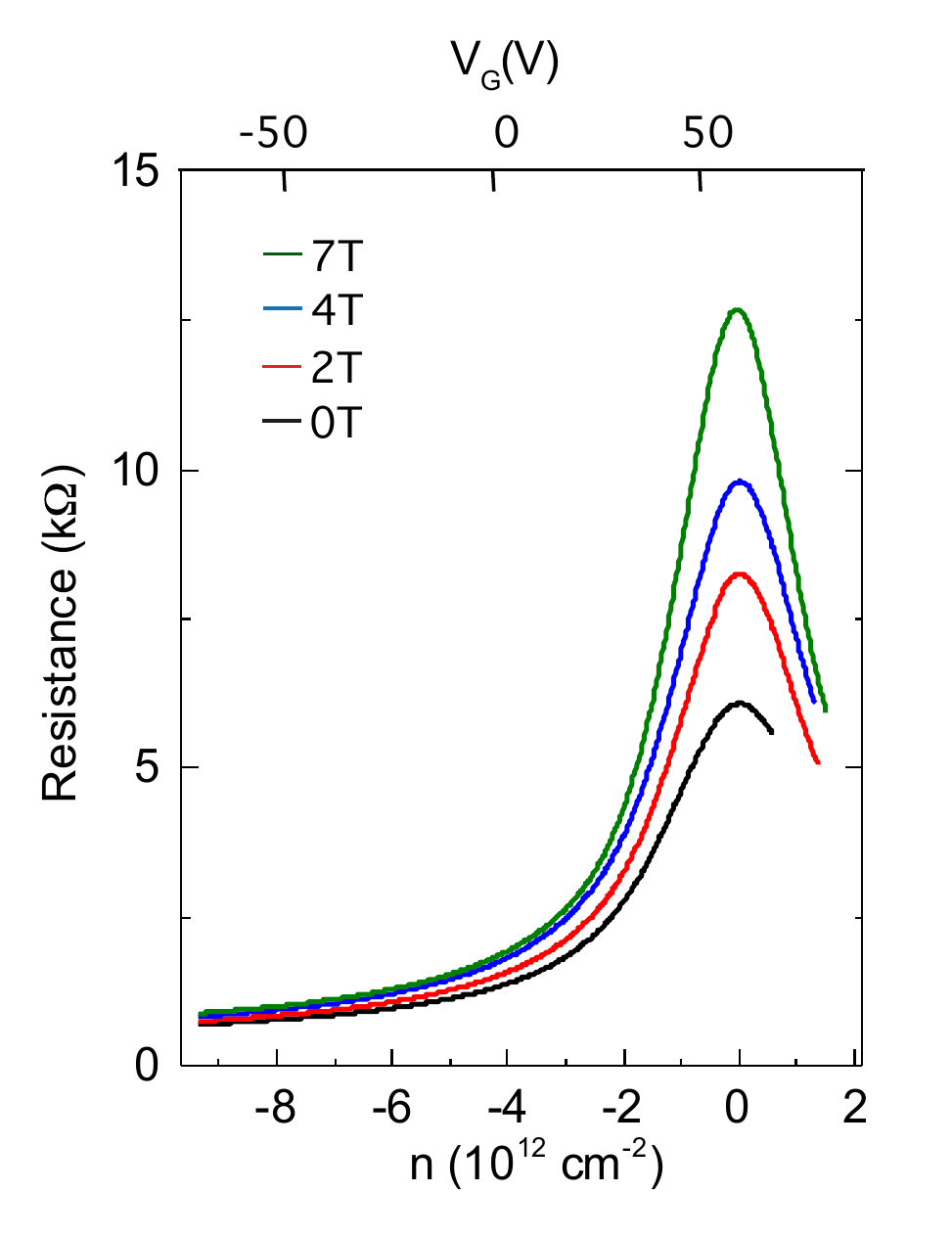}
\caption{\label{FigTransport} \textbf{Supplementary Figure 3.} Magneto-resistivity curves of a patterned g-FET at 250 K, with the same parameters as in Figure 3 of the main text. They are similar to the transport characteristics of continuous graphene (Figure 1 of the main text).}
\end{figure*}

\begin{figure*}[h]
\includegraphics[width=15cm]{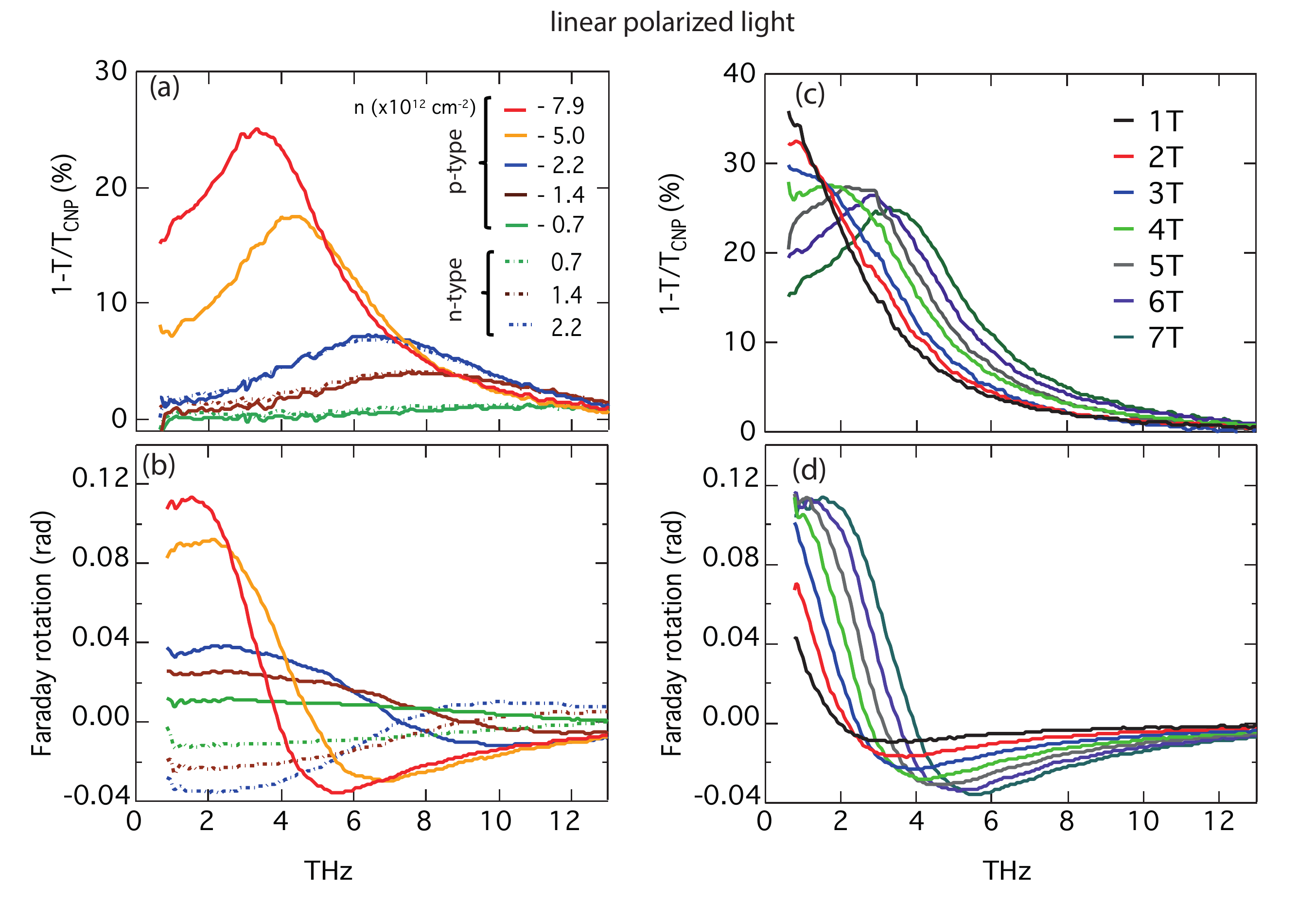}
\caption{\label{FigOriginal} \textbf{Supplementary Figure 4.} The experimental spectra of extinction in linear polarized light and Faraday rotation used to derive the curves on Figures 1c and 1d of the main text. (a) and (b): Linear-polarization extinction (a) and Faraday rotation (b) at B = 7 T for different doping levels. (c) and (d) Linear-polarization extinction (c) and Faraday rotation (d) for $n=-7.9\times10^{12}$ cm$^{-2}$ for different values of magnetic field.}
\end{figure*}

\section{Supplementary Note 1: Model for magneto-optical transmission and Faraday rotation of a single g-FET}

We model a g-FET by a system, where graphene with optical condictivity $\sigma_{\pm}$ (for RH and LH circular polarizations respectively) is deposited on a substrate consisting of two layers (1 and 2) characterized by the complex refractive indices $N_{1}=n_{1} + ik_{1}$ and $N_{2}=n_{2} + ik_{2}$  with the thicknesses $d_{1}$ and $d_{2}$ respectively (Supplementary Figure \ref{FigSetup}). In the present case the first layer is SiO$_{2}$ with $d_{1}=300 $ nm, while the second layer is Si with a thickness of several hundreds of microns. Due to the very small thickness of the first layer, the transmission is mostly determined by the second layer (except in the optical region where SiO$_{2}$ shows a phonon absorption, which is beyond the considered THz range). Nevertheless, for the sake of accuracy we take both layers into account in our calculations.

First we compute all relevant Fresnel coefficients (defined in the figure):
$t_{\text{vg}1}^{\pm}=2/(N_{1}+1+Z_{0}\sigma_{\pm})$, $t_{12}=2N_{1}/(N_{1}+N_{2})$, $t_{21}=2N_{2}/(N_{1}+N_{2})$,  $t_{2\text{v}}=2N_{2}/(N_{2}+1)$, $r_{1\text{gv}}^{\pm}=(N_{1}-1-Z_{0}\sigma_{\pm})/(N_{1}+1+Z_{0}\sigma_{\pm})$, $r_{12}=(N_{1}-N_{2})/(N_{1}+N_{2})$, $r_{2\text{v}}=(N_{2}-1)/(N_{2}+1)$, $r_{21}=(N_{2}-N_{1})/(N_{1}+N_{2})$, $\phi_{1}=\exp(i\omega N_{1}d_{1}/c)$ and $\phi_{2}=\exp(i\omega N_{2}d_{2}/c)$, where $Z_{0}$ is the impedance of vacuum. Note that in the thin-film approximation the effect of graphene can be included into the Fresnel coefficients $t_{\text{vg}1}^{\pm}$ and $r_{1\text{gv}}^{\pm}$.

As the thickness of Si is comparable with or larger than the wavelength, the Fabry-Perot effect is normally produced. However, we suppress the Fabry-Perot effect by reducing the resolution, in order to simplify the data analysis. Therefore we should consider separately the cases of (i) a coherent (amplitude) and (ii) an incoherent (intensity) addition of the multiply-reflected waves in the second layer. Treatment of incoherent addition is described in Supplementary Reference 1.

\emph{Coherent addition}. In this case, we can calculate the total amplitude transmission coefficients:
\begin{equation}\label{TransCoh}
t_{\pm}=\frac{t_{\text{vg}1}^{\pm}t_{12}t_{2\text{v}}\phi_{1}\phi_{2}}{1-r_{1\text{gv}}^{\pm}r_{12}\phi_{1}^2-(r_{21} + r_{1\text{gv}}^{\pm}\phi_{1}^2)r_{2\text{v}}\phi_{2}^2}.
\end{equation}
\noindent The measured intensity transmission coefficients for RH and LH circular polarizations and the Faraday rotation are then obtained straightforwardly: $T_{\pm}=|t_{\pm}|^2$ and $\theta_{\text{F}}=(1/2)\mbox{Arg}(t_{-}t_{+}^{*})$.

\emph{Incoherent addition}. In this case, the phase information in $\phi_{2}$ is lost and intensities are added rather than amplitudes in the second layer. Therefore we have:
\begin{equation}\label{TransIncoh}
T_{\pm}=\frac{\left|t_{\text{vg}1}^{\pm}t_{12}t_{2\text{v}}\phi_{1}\phi_{2}\right|^2}{\left|1-r_{1\text{gv}}^{\pm}r_{12}\phi_{1}^2\right|^2-\left|r_{21} + r_{1\text{gv}}^{\pm}\phi_{1}^2\right|^2 \left|r_{2\text{v}}\phi_{2}^2\right|^2}.
\end{equation}
\begin{equation}\label{FaraIncoh}
\theta_{\text{F}}=\frac{1}{2}\mbox{Arg}\left\{\frac{t_{\text{vg}1}^{-}t_{\text{vg}1}^{+*}}{(1-r_{1\text{gv}}^{-}r_{12}\phi_{1}^2)(1-r_{1\text{gv}}^{+}r_{12}\phi_{1}^2)^{*}-(r_{21} + r_{1\text{gv}}^{-}\phi_{1}^2)(r_{21} + r_{1\text{gv}}^{+}\phi_{1}^2)^{*} \left|r_{2\text{v}}\phi_{2}^2\right|^2}\right\}.
\end{equation}

Expressions for the reflection coefficients and magneto-optical Kerr rotation angle can be derived in a similar fashion.

\section{Supplementary Note 2: Relation between transmission, reflection, absorption and extinction}

A model calculation for a g-FET with a high doping level ($n=-7.2\times 10^{12}$ cm$^{-2}$) at 7 T is shown in Supplementary Figure \ref{FigSpectra}. The top panel presents the model reflectivity and transmission spectra for LH polarized light (where the CR is observed) and at a charge neutrality point, where we assume the Drude weight to be zero. In the calculation, the multiple reflections in the substrate were added incoherently, which corresponds to our experiment, where the Fabry-Perot effect was suppressed. One can see that doping has a much stronger effect on transmission than on reflection. The bottom panel compares the absorption $A= 1 - R - T$, which is dominated by graphene charge carriers, and the extinction spectra  $1 - T/T{\mbox{\tiny CNP}}$, which can be most accurately measured in the experiment. An important conclusion is that under typical conditions of our experiment the extinction has essentially the same spectral shape as the absorption by the Drude carriers, with a strong peak at the CR frequency $\omega_{\text{c}}$. The difference by about a factor of 2 is due to a high refractive index of Si, which makes the transmission at the CNP close to 50\%.

\newpage

\textbf{Supplementary References}

[1] Harbecke, B Coherent and incoherent reflection and transmission of multilayer structures, \emph{Appl. Phys. B} \textbf{39}, 165-170 (1986).